\documentstyle[12pt]{article}
         \makeatletter
         \def\thefigure{\@arabic\c@figure}\def\fps@figure{tbp}
         \def\ftype@figure{1}\def\ext@figure{lof}
         \def\fnum@figure{\protect\footnotesize Fig.\ \thefigure}
         \def\thetable{\@arabic\c@table}
         \def\fps@table{tbp}\def\ftype@table{2}\def\ext@table{lot}
         \def\fnum@table{\protect\footnotesize Table \thetable}
         \makeatother
\begin{document}
\vspace*{0.3in}
\begin{center}
  {\Large \bf 
Comment on\\
``A Quantum-Mechanical Equivalent-Photon Spectrum
for Heavy Ion Physics" 
}\\
  \bigskip
  \bigskip
  {\Large  G. Baur$^a$ and C.A. Bertulani$^{b,}$\footnote{
Permanent Address:
Instituto de F\'\i sica, Universidade Federal do Rio de Janeiro,
   21945-970  Rio de Janeiro,  Rj,    Brazil.
E-mail: bertu@if.ufrj.br. Work partially supported by the
                Conselho Nacional de Desenvolvimento Cient\'\i fico e
                Tecnol\'ogico, CNPq/Brazil.}
 }  \\
  \bigskip

   $^a$ Forschungsanlage J\"ulich, Postfach  1913\\
   D-52425 J\"ulich, Germany\\

   $^b$ 
   Gesellschaft f\"ur Schwerionenforschung, KPII\\
   Planckstr. 1, D-64291 Darmstadt, Germany\\

 \bigskip

  \end{center}
\begin{abstract}
We critically discuss the recently developed quantum-mechanical equivalent
photon spectrum by Benesh {\it et al} 
\cite{Ben96}. We point out that the key point, the strong absorption in heavy ion collisions, is not treated adequately.  Conclusions
drawn from such a spectrum are invalid. Equivalent
photon spectra
appropriate for heavy ions, have been given before in quantal as well as
semiclassical versions and were found to be very satisfactory.
\end{abstract}
\baselineskip 4ex

In a recent paper \cite{Ben96} a quantum mechanical equivalent photon
spectrum for heavy ion physics was calculated. Significant deviations from the prediction of  previous calculations
for mildly relativistic
collisions ($\gamma < 2-3$) were found \cite{Ben96}. This is surprising,
since the usual assumption of classical trajectories in semiclassical
calculations, or eikonal wave functions in quantal calculations, are well
known as valid for heavy ion reactions \cite{WA79,BB88}. 

We can trace the origin of the discrepancy to the inadequate treatment of strong interaction effects in ref. \cite{Ben96}. Their eq. (1) is based
on the plane-wave Born approximation. In the notation of ref. \cite{BB88}
\begin{equation}
T_{Born}= {1\over c} \int d^3 r A_\mu ({\bf r}) <I_fM_f|J_\mu({\bf r})|I_iM_I> \ ,
\end{equation}
where $A_\mu$ represents the four-potential created by the transition 
current of the projectile. Our point is most clearly explained in the
case of extreme strong absorption ({\it black disk} model).
We give a quantum-mechanical equivalent-photon spectrum using Glauber
wave functions for the projectile in the initial and final state 
\cite{BB88,BN93}. Thus the Glauber phase is given by
\begin{eqnarray}
e^{i\chi(b)}&=&0 \ , \ \ \ \ {\rm for} \ \ b<R \ , \nonumber \\
&=& e^{i\psi_c(b)}  \ , \ \ \ \ {\rm for} \ \ b>R \ ,
\end{eqnarray}
where $R$ denotes the strong absorption radius. The Glauber Coulomb phase
is denoted by $\psi_c(b)$.
The appropriate $T$-matrix is now given by \cite{BB88,BN93}
\begin{equation}
T_{fi}({\bf q})= {1\over {(2\pi)^2}} 
\int_{b>R} d^2 b \ e^{i{\bf q_\perp .b}} \
e^{i\psi_c(b)} \int d^2q'_\perp e^{-i{\bf q'_\perp.b}}
\ T_{Born} ({\bf q'_\perp} , q_{//})\ .
\end{equation}
This leads to an equivalent photon spectrum (see eq. (12) of Ref. \cite{BN93}).
In this derivation, it was used that the projectile and target do not overlap 
\cite{WA79}. 
The form factor of the projectile charge distribution does not
enter, since the electric field of a spherically symmetric charge distribution
depends only on the charge contained within its radius $R$.
Diffraction effects due to the finite wavelength of the
projectile are taken into account in this approach (see e.g. figure 3 of
ref. \cite{BN93}, or figs. 2-4 of ref. \cite{BR96}). It is interesting to note that
the total (i.e. angle integrated) cross-section is the same for the semiclassical
and quantum treatment in the sharp-cutoff model \cite{BB88,BN93}.

It is evident from eq. (2) that an adequate treatment of
strong interaction effects cannot be obtained by using a Born-approximation
$T$-matrix and calculate
total cross sections by introducing
a phenomenological  cutoff $q_{max}\sim 1/R$ on the
transverse momentum transfer ${\bf q}_\perp$, as it was done in
ref. \cite{Ben96}. At most, this would lead to a very approximate
result. A good quantitative description
of the cross sections, as stated by the authors,
cannot be obtained.  The key point
here is that this calculation is better treated in coordinate space, since
the strong absorption is treated in a simple way. In momentum 
space one has
to introduce momentum cutoffs which do not have a one-to-one correspondence
with r-space (or impact parameter space) cutoffs.

Moreover, in ref. \cite{Ben96} it was stated that ``the new spectra ... leave
little room for more exotic multiphonon mechanisms required in a 
semiclassical analysis". Such criticism of refs. \cite{NB93,LB93}
(refs. 20 and 21 of ref. \cite{Ben96}) is based on a wrong assumption and
is therefore invalid. Indeed, multiphonon effects are a natural consequence
of QED and have been clearly observed in relativistic electromagnetic 
excitation \cite{Eml94,Bor96}.

A small effect worth mentioning in this context is the 
change of the state of the
projectile during the excitation process. It is of a genuinely quantum mechanical
nature. This leads to mutual excitation, and a simple
form of the corresponding equivalent photon
spectrum is given in ref. \cite{Hen96}. The case of a deformed projectile is
treated in ref. \cite{Ber93}.

\end{document}